\documentclass[conference]{IEEEtran}
\IEEEoverridecommandlockouts

\usepackage[noadjust]{cite}
\usepackage{amsmath,amssymb,amsfonts}
\usepackage{algorithmic}
\usepackage{graphicx}
\usepackage{textcomp}
\usepackage{xcolor}
\usepackage{hyperref}
\def\BibTeX{{\rm B\kern-.05em{\sc i\kern-.025em b}\kern-.08em
    T\kern-.1667em\lower.7ex\hbox{E}\kern-.125emX}}
\begin{document}

\title{	ListenToJESD204B: A Lightweight Open-Source JESD204B IP Core for FPGA-Based Ultrasound Acquisition systems 
\thanks{The authors acknowledge support from the ETH Research Grant \mbox{ETH-C-01-21-2} (Project ListenToLight).}
}

\author{
\IEEEauthorblockN{
Soumyo~Bhattacharjee\IEEEauthorrefmark{1},
Federico~Villani\IEEEauthorrefmark{1},
Christian~Vogt\IEEEauthorrefmark{1},
Andrea~Cossettini\IEEEauthorrefmark{1},
Luca~Benini\IEEEauthorrefmark{1}}
\IEEEauthorblockA{Email: {sbhattacharj@student.ethz.ch, villanif@ethz.ch} \\
    \IEEEauthorrefmark{1}Department of Information Technology and Electrical Engineering, ETH Zurich, Switzerland\\
}%
}
\IEEEoverridecommandlockouts
\IEEEpubid{\makebox[\columnwidth]{979-8-3315-6578-7/25/\$31.00 \copyright2025 IEEE~\hfill} \hspace{\columnsep}\makebox[\columnwidth]{ }}
\maketitle
\IEEEpubidadjcol
\begin{abstract}

The demand for hundreds of tightly synchronized channels operating at tens of MSPS in ultrasound systems exceeds conventional low-voltage differential signaling links' bandwidth, pin count, and latency. Although the JESD204B serial interface mitigates these limitations, commercial FPGA IP cores are proprietary, costly, and resource-intensive. We present ListenToJESD204B, an open-source receiver IP core released under a permissive Solderpad 0.51 license for AMD Xilinx Zynq UltraScale+ devices. Written in synthesizable SystemVerilog, the core supports four GTH/GTY lanes at 12.8 Gb/s and provides cycle-accurate AXI-Stream data alongside deterministic Subclass~1 latency. It occupies only 107 configurable logic blocks (approximately 437 LUTs), representing a 79\% reduction compared to comparable commercially available IP. A modular data path featuring per-lane elastic buffers, SYSREF-locked LMFC generation, and optional LFSR descrambling facilitates scaling to high lane counts.
We verified protocol compliance through simulation against the Xilinx JESD204C IP in JESD204B mode and on hardware using TI AFE58JD48 ADCs. Block stability was verified by streaming 80 MSPS, 16-bit samples over two 12.8 Gb/s links for 30 minutes with no errors.

\end{abstract}
\makeatletter
\def\ps@IEEEtitlepagestyle{
  \def\@oddfoot{\mycopyrightnotice}
  \def\@evenfoot{}
}
\def\mycopyrightnotice{
  \begin{minipage}{\textwidth}
  \centering \scriptsize
  Copyright~\copyright~2023 IEEE. Personal use of this material is permitted. Permission from IEEE must be obtained for all other uses, in any current or future media, including\\reprinting/republishing this material for advertising or promotional purposes, creating new collective works, for resale or redistribution to servers or lists, or reuse of any copyrighted component of this work in other works.
  \end{minipage}
}
\makeatother


\section{Introduction}

FPGA-based ultrasound systems enable flexible, real-time diagnostic imaging for critical medical applications, including emergency point-of-care scenarios, wearable health monitoring, and rapid bedside diagnostics, facilitating immediate clinical decision-making and improved patient outcomes~\cite{schellenbergHandheldOptoacousticImaging2018}. High-speed data acquisition is fundamental to these imaging technologies, requiring efficient and reliable data transfer from analog front-end (AFE) devices to digital processing units~\cite{boniUltrasoundOpenPlatforms2018}. As imaging systems scale toward higher channel counts, increased sampling rates, and compact form factors, traditional interfaces like Low Voltage Differential Signaling (LVDS) become limiting due to limited scalability, high pin counts, and signal integrity issues at high data rates~\cite{boniULAOP256256Channel2016,jensenSARUSSyntheticAperture2013}. Such limitations particularly hinder emerging portable biomedical systems such as TinyProbe~\cite{vostrikovTinyProbeWearable32Channel2025} and wearable imaging platforms~\cite{speicherWearable256ElementMUXBased2025}, which rely on compact FPGAs with limited Input Output (IO) interfaces.

The JESD204B standard provides a scalable, standardized alternative, reducing interface complexity and ensuring deterministic latency, which is critical for accurate real-time imaging and beamforming. Existing commercial IP solutions, such as Xilinx LogiCORE JESD204 IP (v5.2), are mature but suffer from restrictive licensing models, limited transparency, and high costs, limiting their applicability in open research and modular platform development. In addition, such commercial solutions need to support a large number of alternative configurations, thereby resulting in sub-optimal FPGA resource utilization when a minimal and lightweight implementation is desired, such as for wearable systems. Hence, despite the clear advantages over LVDS, existing solutions and open-source adoption of JESD204B remain limited by these bottlenecks.

To address this gap, we present \textit{ListenToJESD204B}, a fully open-source, resource-efficient JESD204B Subclass~1 receive-only IP core written in synthesizable SystemVerilog. It is targeted at the AMD Zynq UltraScale+ ZU19EG MPSoC and implements the complete JESD204B Subclass~1 receive specification, supporting four lanes at up to 12.8 Gbps each.

Our key contributions are:
\begin{itemize}
  \item \textbf{Open-source Subclass~1 RX IP:} A publicly available SystemVerilog core implementing the full JESD204B receive stack at maximum standard bit-rates.
  \item \textbf{Resource-efficient architecture:} Demonstration of 79\% reduction in Look-Up Table (LUT) and Flip-Flop(FF) utilization compared to a leading commercial IP, validated via both simulation and silicon on the ZU19EG MPSoC.
  \item \textbf{High-speed ultrasound capture:} Demonstration of reliable acquisition of 80~MSps analog signals over two 12.8~Gbps JESD204B lanes through an analog front-end.
\end{itemize}

\section{Background and Related Work}

FPGA-based architectures are commonly used in biomedical ultrasound imaging due to their intrinsic parallelism, deterministic timing, and reconfigurability. Early FPGA implementations using LVDS were suitable only for modest channel counts~\cite{gi-duckkimSingleFPGAbasedPortable2012}. Open platforms by Boni et al.~\cite{boniUltrasoundOpenPlatforms2018} emphasized rapid development and experimentation but remained constrained by LVDS limitations, particularly affecting scalability and synchronization in high-channel-count scenarios.

Systems such as Risser et al.'s 1024-channel volumetric ultrasound platform~\cite{risserRealTimeVolumetricUltrasound2021} further illustrated the complexity of managing multi-FPGA setups since traditional methods of interfacing with AFEs required a very high pin count for the ever-growing channel count.

Since its introduction by JEDEC~\cite{jedecJESD204BSerialInterface2011}, JESD204B has become preferred for high-speed ADC interfacing due to scalability and deterministic latency. However, widespread adoption within open-source communities remains limited due to dependence on proprietary IP cores, as illustrated by Wang et al.~\cite{wangHighSpeedADCInterface2023}. Even when certain implementations have used the JESD204B standard, the source code of the design is not openly available~\cite{gonzalezHardwareAccelerationDigital2024}.

Current open-source JESD204B solutions like Analog Devices' HDL framework~\cite{analogdevicesJESD204InterfaceHDL} and LiteJESD204B~\cite{enjoy-digitalLiteJESD204BOpenSourceJESD204B2025} are important references but still suffer from some issues. The former implementation is not permissively licensed and requires modifications to make the design operate correctly on AMD Xilinx Ultrascale+ platforms with AFEs of different vendors, while the latter is based on Migen~\cite{m-labsMigen2025}, a toolchain that is not widely adopted, has only been verified at line rates up to 6.25 Gbps and does not support not scrambled mode.

\textit{ListenToJESD204B addresses the critical gaps by providing a compact, easy-to-integrate and extend, and a fully open-source IP core}. Supporting AMD Xilinx JESD204 Physical layer (PHY) block compatibility, deterministic Subclass~1 latency, and robust infrastructure, it bridges the gap toward scalable, reproducible, and community-driven biomedical ultrasound system design research.

\section{System Context}

\begin{figure}
  \centering
  \includegraphics[width=0.9\columnwidth]{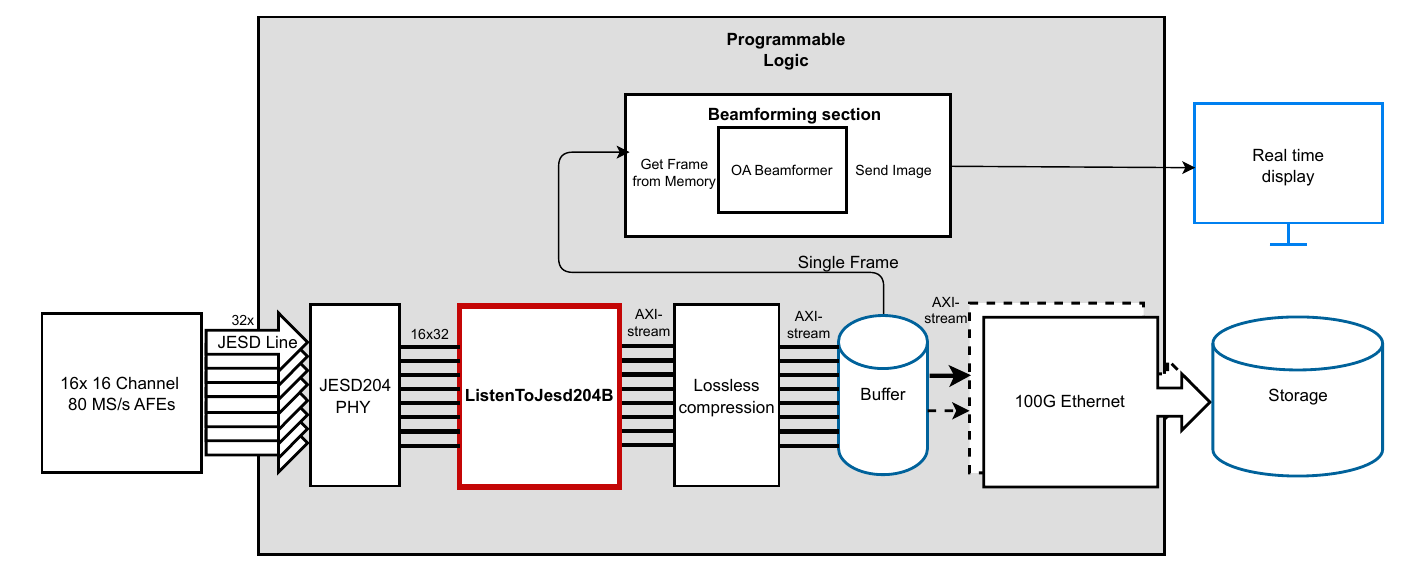}
  \caption{Block diagram of a 256-channel open-ultrasound front-end.  
 The \textit{ListenToJESD204B} core converts 32 JESD204B lanes to parallel data, which is then losslessly compressed and stored, as well as reconstructed for real-time imaging.}
  \label{fig:system_context}
\end{figure}

Fig.~\ref{fig:system_context} illustrates the integration of the \textit{ListenToJESD204B} receiver into a compact, high-performance open-ultrasound front-end architecture optimized for state-of-the-art imaging. Sixteen TI AFE58JD48 analog front-ends digitize 16 channels each at 80 MSamples/s and 16-bit resolution, producing approximately 327~Gb/s of data across 256 channels. Each AFE streams data via a dual-lane JESD204B interface, resulting in 32 high-speed lanes interfacing directly with the GTH transceivers of a AMD Xilinx ZU19EG System on Chip (SoC). The vendor-provided JESD204 PHY combined with the open-source \mbox{\textit{ListenToJESD204B}} receiver performs deserialization and word alignment, delivering a 256-bit parallel stream to the FPGA fabric.

Due to the impracticality of DRAM storage at these data rates, the data stream undergoes immediate real-time compression~\cite{villaniFPGAAcceleratedHybridLossless2024}. Ultrasound echo data typically achieves compression factors of 4--6, reducing the required outbound bandwidth to approximately 70~Gb/s. This comfortably fits into one RDMA-capable 100 Gb Ethernet link~\cite{cossettiniRDMAInterfaceUltraFast2022}, facilitating direct forwarding to GPU clusters or local NVMe archival. The compact architecture of the data path allows a reconfigurable block to be integrated into the FPGA fabric that can perform real-time imaging and beamforming~\cite{villaniAdaptiveImageReconstruction2024}, enabling immediate visualization of the acquired data.

\vfill\break
\section{Architecture and Implementation}

The \textit{ListenToJESD204B} Rx IP implements a modular, simple, and synthesizable SystemVerilog design for a JESD204B Subclass~1 interface on FPGA platforms. Its architecture comprises protocol-specific sub-modules connected by standardized streaming interfaces, facilitating reuse, independent verification, and integration into broader acquisition pipelines.

\begin{figure}[htbp]
    \centering
    \includegraphics[width=0.9\linewidth]{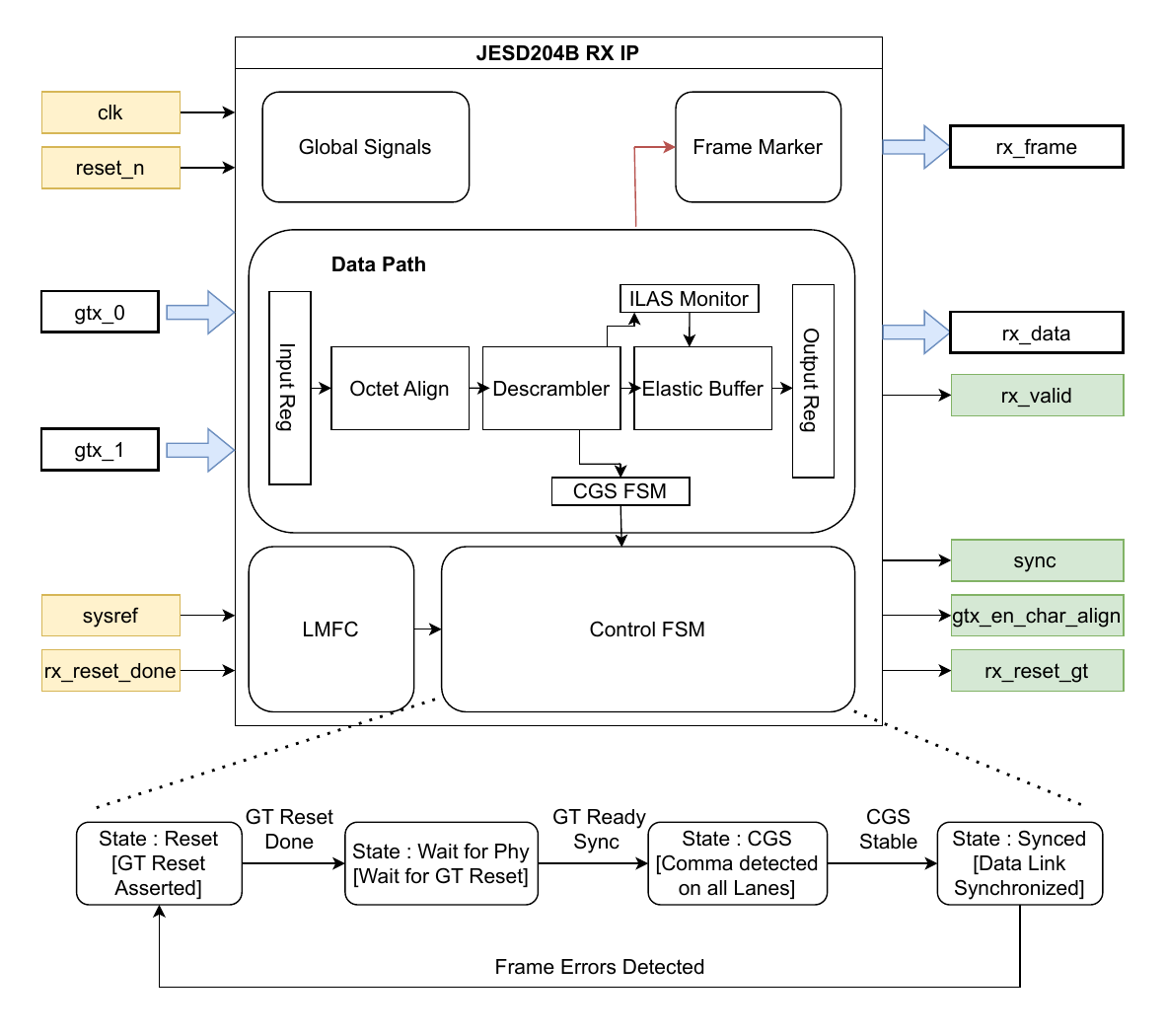}
    \caption{High-Level Architecture of the JESD204B Receiver}
    \label{fig:architecture}
\end{figure}

Designed to interface with Xilinx's JESD204 PHY using GTY/GTH transceivers, the IP supports up to four lanes and line rates from 6 Gbps to 12.8 Gbps. It outputs AXI-Stream-compatible data, simplifying downstream connections. To make the IP user-friendly, the maximum number of lanes is set to 4 per link, but can be easily modified at the wrapper level. All underlying modules are compatible with up to 32 lanes per link. Fig.~\ref{fig:architecture} illustrates the high-level architecture of the JESD204B receiver, and Table~\ref{tab:parameters} lists the IP configurable parameters.

\begin{table}[htbp]
\centering
\caption{Configurable Parameters of the Receiver IP}
\label{tab:parameters}
\begin{tabular}{|l|l|l|}
\hline
\textbf{Parameter} & \textbf{Description} &\textbf{Range} \\
\hline
\texttt{links} & Number of JESD204B links supported & up to 4\\
\texttt{L} & Number of lanes per link & up to 4\\
\texttt{F} & Octets per frame & 4 - 32 \\
\texttt{K} & Frames per multiframe & 1 - 32\\
\texttt{DESCRAMBLING} & Enable flag for data descrambling  & 0 / 1\\
\texttt{DATA\_WIDTH} & Width of the datapath (in bits) & 32 bits\\
\hline
\end{tabular}
\end{table}

\subsection{Data Path}

Each JESD204B lane connects to a dedicated \texttt{data\_path} module responsible for processing four 8-bit octets per cycle. These modules register input data, perform Code Group Synchronization (CGS), validate the Initial Lane Alignment Sequence (ILAS), execute optional descrambling, and align data using per-lane elastic buffers. By encapsulating CGS and ILAS logic within the datapath, the architecture avoids coupling between control and data movement, maintaining clean streaming semantics.

The top-level receiver aggregates lane outputs and emits AXI-Stream-compatible data with a \texttt{rx\_valid} signal. The core assumes continuous data consumption by the downstream system and does not support backpressure via \texttt{tready}. GT configuration, including PLL lock and 8b/10b decoding, is expected to be managed externally, though hooks are provided for tighter integration.

\subsection{Control State Machine}

A finite state machine (FSM) coordinates link initialization across five main states: \texttt{ST\_RESET}, \texttt{ST\_WAIT\_FOR\_PHY}, \texttt{ST\_CGS}, \texttt{ST\_ILAS}, and \texttt{ST\_SYNCED}. In \texttt{ST\_CGS}, the FSM monitors comma-aligned \texttt{/K28.5/} characters and asserts the JESD204B \texttt{SYNC} signal to prompt transmitter alignment. Upon consistent CGS detection, the FSM advances to ILAS verification, checking all multiframes for valid configuration fields. Once lane alignment is confirmed, the system transitions to data reception.

A stability flag and cycle counter gate state transitions to prevent premature advancement due to transients. Fault detection logic with configurable thresholds ensures re-entry to CGS in the event of misalignment.

\subsection{Octet Alignment and Descrambling}

Incoming 10-bit symbols from the transceivers are first decoded and byte-aligned using the \texttt{octet\_align} module, which detects the \texttt{/K28.5/} character and rotates incoming words accordingly. This ensures proper framing of subsequent ILAS and data octets. Descrambling is performed using a pipelined 32-bit-wide Linear Feedback Shift Register(LFSR)-based module implementing $ G(x) = x^{14} + x^{13} + 1 $. Both functions operate independently per lane and are configurable via RTL parameters.

\subsection{SYSREF-Synchronized LMFC Generation}

To enforce deterministic latency, the receiver includes a SYSREF-synchronized LMFC generator. A counter resets on SYSREF edges and rolls over every $ F \times K $ octets, establishing repeatable multiframe boundaries. This synchronized LMFC is used to control elastic buffer release across lanes.

\subsection{Elastic Buffer}

Per-lane elastic buffers implemented as circular FIFOs compensate for skew and align data to LMFC boundaries. A centralized \texttt{buffer\_release} module monitors readiness across all buffers and triggers simultaneous data release. Buffer depth is statically configurable to accommodate application-specific skew tolerances.

\subsection{Design Considerations}

The IP was developed with portability and integration in mind. Key design choices include:
\begin{itemize}
    \item \textbf{Unified Clocking:} All internal logic is driven by a single clock to avoid clock domain crossing(CDC) issues.
    \item \textbf{Fixed Data Width:} 32-bit-wide processing aligns with transceiver output and AXI-Stream expectations.
    \item \textbf{Compact Control Logic:} FSMs are shallow and optimized for timing closure at 320~MHz.
    \item \textbf{Modular Layout:} Clear separation of datapath, control, and synchronization aids testability.
    \item \textbf{Vivado IP Packager Integration:} GUI configuration and block design compatibility support ease of adoption.
\end{itemize}

\begin{figure}[htbp]
    \centering
    \includegraphics[width=0.9\linewidth]{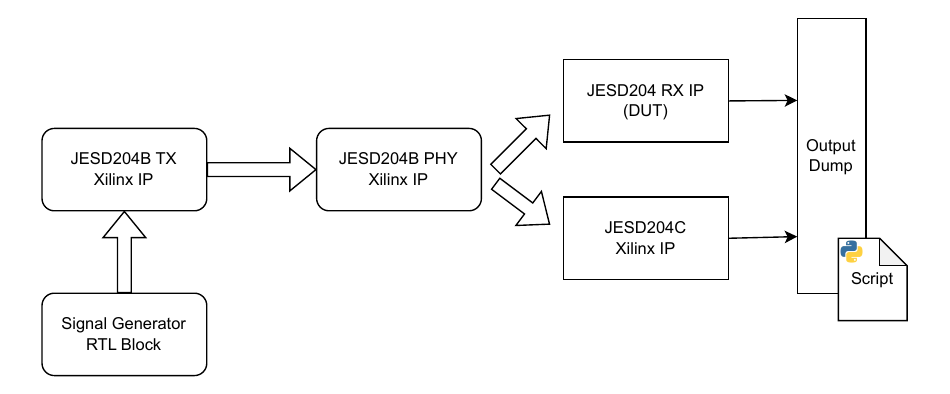}
    \caption{Simulation-based verification setup for the JESD204B receiver IP core. The testbench uses a golden output to evaluate the functional correctness of the implemented design }
    \label{fig:sim}
\end{figure}
\section{Verification methodology and Results}

\begin{figure*}[htbp]
    \centering
    \begin{minipage}{0.32\textwidth}
        \centering
        \includegraphics[width=\textwidth]{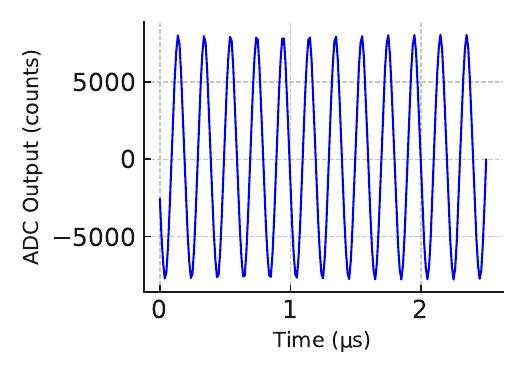}
        \caption{Acquisition and reconstruction of a 5MHz sine wave using the JESD204 RX IP}
        \label{fig:result_01}
    \end{minipage}
    \hfill
    \begin{minipage}{0.64\textwidth}
        \centering
        \includegraphics[width=\textwidth]{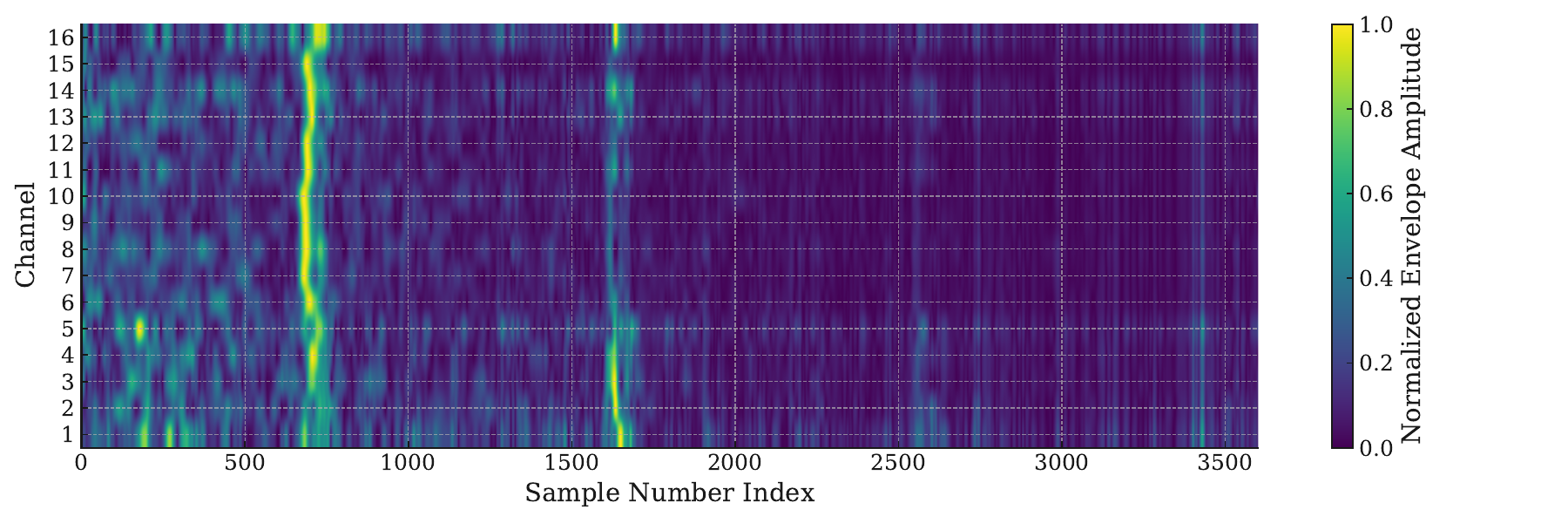}
        \caption{Acquisition of scatterers on CIRS 040GSE Ultrasound Phantom}
        \label{fig:result_02}
    \end{minipage}
\end{figure*}

A two-stage methodology was used to evaluate functional correctness and practical integration of the \textit{ListenToJESD204B} receiver: behavioral simulation and hardware-in-the-loop testing. This confirmed protocol compliance, Subclass~1 deterministic behavior, and compatibility with real-time data acquisition systems.

\subsection{Simulation-Based Verification}

Initial verification was performed in a simulation environment derived from Xilinx's JESD204C example design. The testbench (see Fig.~\ref{fig:sim}) was modified to insert the receiver IP as the device under test (DUT) and configured to emulate a JESD204B transmitter using Xilinx JESD204C IP in B-mode. Octets per frame and frames per multiframe were matched between both modules to ensure consistency.

Simulations were executed in Vivado 2023.2. Test vectors included CGS, ILAS, and randomized payloads to verify the handling of typical and edge-case traffic patterns. Receiver outputs were compared against a reference implementation configured identically. Assertions validated proper transition through FSM states, ILAS extraction, and multiframe alignment.

\subsection{Hardware Evaluation Setup}

\begin{figure}[htbp]
    \centering
    \includegraphics[width=1\linewidth]{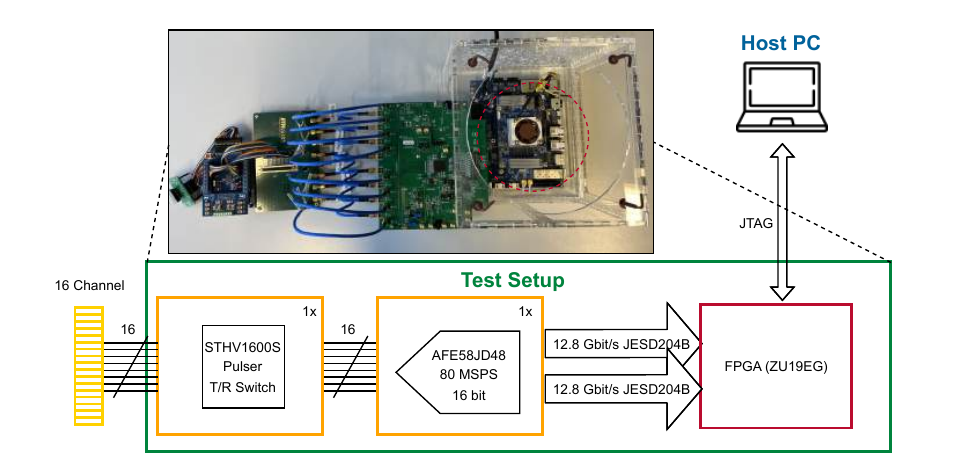}
    \caption{Experimental test setup used for validating the JESD204B RX core}
    \label{fig:hardware_setup}
\end{figure}

To validate real-time performance, the proposed receiver IP was deployed on a Xilinx Zynq UltraScale+ MPSoC (ZU19EG) as part of a complete ultrafast ultrasound acquisition platform (shown in Fig.~\ref{fig:hardware_setup}) featuring a 5MHz 32 channel ultrasound transducer (Vermon), of which 16 channels were connected to a 16 channel STHV1600 ultrasonic pulser (STMicroelectronics), a 16 channel, 80 MSamples/s AFE58JD48 analog front-end (AFE) (Texas Instruments), streaming in 2 JESD link, 12.8Gsamples/s to the ZU19EG FPGA. The SYSREF signal required for Subclass~1 synchronization and clocking was provided by an LMK04826B clock conditioner (Texas Instruments). Data received by the JESD204B core was captured utilizing an Integrated Logic Analyser buffer and transferred to a host PC via JTAG. Additionally, internal signals, including the LMFC clock, FSM states, and data alignment flags, were monitored using ILA. To test for long-term stability, experiments were run for thirty minutes. During this time, no invalid frames were observed. This configuration assessed acquisition stability, link synchronization, and protocol compliance under realistic operating conditions, slightly over the 12.5Gbps limit of the JESD204B specifications.

Multiple experiments were conducted to validate integration with real-world data converters and external signal sources. In the first test, the ADC transmitted a digital ramp pattern across 16 channels at 80 MSPS, interleaved into two lanes operating at 12.8 Gbps. The receiver core successfully aligned the 8B/10B-encoded stream, performed descrambling, and forwarded reconstructed samples to a high-speed buffer for analysis (not shown). Lane synchronization was achieved within 15 frame clock cycles from initialization.

In the second test, a precision 5 MHz sine wave (0.6Vpp, 0V offset) from a signal generator was connected to the AFE58JD48 input channels. The receiver accurately reconstructed the waveform in both scrambled and descrambled modes. Fig. \ref{fig:result_01} shows the reconstructed signal.

The third test used a Cirs 040GSE (Sun Nuclear) tissue-mimicking phantom. The acquisition ultrasound transducer was positioned to image the vertical distance scatterers of the phantom (equidistant reflectors spaced at 1.5 cm intervals). The transducer was excited with a 3-cycle, 5 MHz, 20VPP Plane Wave 0 pulse to generate the ultrasound signals. The receiver captured 8000 samples across all 16 channels at 80 MSPS, which were subsequently bandpass filtered (1-10 MHz), envelope-detected using a Hilbert transform, and windowed. Fig. \ref{fig:result_02} shows the resulting raw channel data, which reveals four distinct reflectors at even spacing and validates the design on a realistic end-to-end ultrasound acquisition pipeline.

\subsection{Timing and Resource Utilization}

The receiver was synthesized using Vivado 2023.2, targeting the Zynq UltraScale+ ZU19EG device. The core met timing at a 320~MHz clock rate (Maximum Clock supported by the JEDEC JESD204B standard) with a worst negative slack of 0.31~ns.

\begin{table}[ht]
\centering
\caption{Resource Utilization Comparison for a Two-Channel Implementation}
\label{tab:utilization}
\begin{tabular}{lccc}
\hline
\textbf{Resource} & \multicolumn{2}{c}{\textbf{This work}}  & \textbf{Xilinx IP}  \\
 & \textbf{Descrambling} & \textbf{Descrambling Off} & \textbf{JESD204C}\\
\hline
CLB LUTs         & 437   & 426   &2092 \\
CLB Registers    & 429   & 410   &1988 \\
CARRY8 Blocks    & 2     & 2     &46 \\
F7\&F8 Muxes         & 0     & 0     &96 \\       
CLBs             & 107   & 102   &470 \\
\hline
\end{tabular}
\end{table}

Table~\ref{tab:utilization} reports the resource utilization of our solution. The implementation achieves significant logic savings compared to the Xilinx IP (our implementation utilizes only 21\% of LUTs and registers compared to the Xilinx IP)  while retaining protocol compliance and Subclass~1 determinism. The Xilinx IP additionally offers more reconfigurability with run-time configuration over the AXI interface. Excluding such extra features leaves more space for compute-intensive processing blocks. By minimizing the footprint of the receiver interface, our design makes it feasible to support a higher number of acquisition channels while reserving sufficient FPGA capacity for downstream processing.

A fixed startup latency of 13 clock cycles (40ns) is observed before the first valid sample appears, after which the throughput matches the Xilinx JESD204C IP.

\section{Conclusion}

This paper presented \textit{ListenToJESD204B}, a compact and fully open-source receiver (Rx) IP core that implements the JESD204B Subclass~1 protocol in a modular and resource-efficient manner. Designed in synthesizable SystemVerilog (Packaged using Vivado IP Packager) and tested on Xilinx UltraScale+ FPGA, the core supports multi-lane configurations up to 4 lanes and 12.8~Gbps and integrates cleanly into open hardware platforms. Functional validation was performed through simulation and hardware-in-the-loop testing, demonstrating protocol compliance and stable timing behavior. The design achieves timing closure with a 79\% lower logic utilization compared to a leading commercial IP, making it suitable for embedded FPGA-based systems requiring deterministic high-speed serial reception.

Possible extensions for this work include support for Subclass-0 and Subclass-2 modes and the development of a complementary transmitter IP core to enable full link validation. Another important addition is a modular, reusable testbench framework to facilitate systematic verification of both the current receiver core and future extensions.

\section{Data Availability}
The Open source \textit{ListenToJESD204B} RX IP core is released under the permissive Solderpad 0.51 license and publicly available at: \href{https://github.com/pulp-bio/ListenToJESD}{\texttt{github.com/pulp-bio/ListenToJESD}}

\bibliographystyle{IEEEtran}
\bibliography{OpenJESD.bib}

\begin{thebibliography}{10}
\providecommand{\url}[1]{#1}
\csname url@samestyle\endcsname
\providecommand{\newblock}{\relax}
\providecommand{\bibinfo}[2]{#2}
\providecommand{\BIBentrySTDinterwordspacing}{\spaceskip=0pt\relax}
\providecommand{\BIBentryALTinterwordstretchfactor}{4}
\providecommand{\BIBentryALTinterwordspacing}{\spaceskip=\fontdimen2\font plus
\BIBentryALTinterwordstretchfactor\fontdimen3\font minus \fontdimen4\font\relax}
\providecommand{\BIBforeignlanguage}[2]{{%
\expandafter\ifx\csname l@#1\endcsname\relax
\typeout{** WARNING: IEEEtran.bst: No hyphenation pattern has been}%
\typeout{** loaded for the language `#1'. Using the pattern for}%
\typeout{** the default language instead.}%
\else
\language=\csname l@#1\endcsname
\fi
#2}}
\providecommand{\BIBdecl}{\relax}
\BIBdecl

\bibitem{schellenbergHandheldOptoacousticImaging2018}
M.~W. Schellenberg and H.~K. Hunt, ``Hand-held optoacoustic imaging: {{A}} review,'' \emph{Photoacoustics}, vol.~11, pp. 14--27, Sep. 2018.

\bibitem{boniUltrasoundOpenPlatforms2018}
E.~Boni, A.~C.~H. Yu, S.~Freear, J.~A. Jensen, and P.~Tortoli, ``Ultrasound {{Open Platforms}} for {{Next-Generation Imaging Technique Development}},'' \emph{IEEE Transactions on Ultrasonics, Ferroelectrics, and Frequency Control}, vol.~65, no.~7, pp. 1078--1092, Jul. 2018.

\bibitem{boniULAOP256256Channel2016}
E.~Boni, L.~Bassi, A.~Dallai, F.~Guidi, V.~Meacci, A.~Ramalli, S.~Ricci, and P.~Tortoli, ``{{ULA-OP}} 256: {{A}} 256-{{Channel Open Scanner}} for {{Development}} and {{Real-Time Implementation}} of {{New Ultrasound Methods}},'' \emph{IEEE Transactions on Ultrasonics, Ferroelectrics, and Frequency Control}, vol.~63, no.~10, pp. 1488--1495, Oct. 2016.

\bibitem{jensenSARUSSyntheticAperture2013}
J.~A. Jensen, H.~{Holten-Lund}, R.~T. Nilsson, M.~Hansen, U.~D. Larsen, R.~P. Domsten, B.~G. Tomov, M.~B. Stuart, S.~I. Nikolov, M.~J. Pihl, Y.~Du, J.~H. Rasmussen, and M.~F. Rasmussen, ``{{SARUS}}: {{A}} synthetic aperture real-time ultrasound system,'' \emph{IEEE Transactions on Ultrasonics, Ferroelectrics, and Frequency Control}, vol.~60, no.~9, pp. 1838--1852, Sep. 2013.

\bibitem{vostrikovTinyProbeWearable32Channel2025}
S.~Vostrikov, J.~Tille, L.~Benini, and A.~Cossettini, ``{{TinyProbe}}: {{A Wearable}} 32-{{Channel Multimodal Wireless Ultrasound Probe}},'' \emph{IEEE Transactions on Ultrasonics, Ferroelectrics, and Frequency Control}, vol.~72, no.~1, pp. 64--76, Jan. 2025.

\bibitem{speicherWearable256ElementMUXBased2025}
D.~Speicher, T.~Gr{\"u}n, S.~Weber, H.~Hewener, S.~Klesy, S.~Rumanus, H.~Strohm, O.~Stamm, L.~Perotti, S.~H. Tretbar, and M.~Fournelle, ``Wearable 256-{{Element MUX-Based Linear Array Transducer}} for {{Monitoring}} of {{Deep Abdominal Muscles}},'' \emph{Applied Sciences}, vol.~15, no.~7, p. 3600, Mar. 2025.

\bibitem{gi-duckkimSingleFPGAbasedPortable2012}
{Gi-Duck Kim}, C.~Yoon, {Sang-Bum Kye}, Y.~Lee, J.~Kang, Y.~Yoo, and {Tai-kyong Song}, ``A single {{FPGA-based}} portable ultrasound imaging system for point-of-care applications,'' \emph{IEEE Transactions on Ultrasonics, Ferroelectrics and Frequency Control}, vol.~59, no.~7, pp. 1386--1394, Jul. 2012.

\bibitem{risserRealTimeVolumetricUltrasound2021}
C.~Risser, H.~Hewener, M.~Fournelle, H.~Fonfara, S.~{Barry-Hummel}, S.~Weber, D.~Speicher, and S.~Tretbar, ``Real-{{Time Volumetric Ultrasound Research Platform}} with 1024 {{Parallel Transmit}} and {{Receive Channels}},'' \emph{Applied Sciences}, vol.~11, no.~13, p. 5795, Jun. 2021.

\bibitem{jedecJESD204BSerialInterface2011}
JEDEC, ``{{JESD204B}}: {{Serial Interface}} for {{Data Converters}},'' https://www.jedec.org/sites/default/files/docs/JESD204B.pdf, Jul. 2011.

\bibitem{wangHighSpeedADCInterface2023}
R.~Wang, J.~Zhang, C.~Jiang, and H.~Liu, ``High-{{Speed ADC Interface Design Based}} on {{JESD204B Protocol}},'' in \emph{{{SPIE Smart Photonic}} and {{Electronic Microsystems Conference}} ({{SPCS}})}, 2023.

\bibitem{gonzalezHardwareAccelerationDigital2024}
C.~Gonz{\'a}lez, M.~Ruiz, A.~Carpe{\~n}o, A.~Pi{\~n}as, D.~{Cano-Ott}, J.~Plaza, T.~Martinez, and D.~Villamarin, ``Hardware {{Acceleration}} of {{Digital Pulse Shape Analysis Using FPGAs}},'' \emph{Sensors}, vol.~24, no.~9, p. 2724, Apr. 2024.

\bibitem{analogdevicesJESD204InterfaceHDL}
{Analog Devices}, ``{{JESD204 Interface HDL Framework}},'' https://github.com/analogdevicesinc/hdl/tree/main/library/jesd204.

\bibitem{enjoy-digitalLiteJESD204BOpenSourceJESD204B2025}
{enjoy-digital}, ``{{LiteJESD204B}}: {{Open-Source JESD204B IP Core}},'' https://github.com/enjoy-digital/litejesd204b, Apr. 2025.

\bibitem{m-labsMigen2025}
{m-labs}, ``Migen,'' https://github.com/m-labs/migen, May 2025.

\bibitem{villaniFPGAAcceleratedHybridLossless2024}
F.~Villani, S.~Mathys, {\c C}.~{\"O}zsoy, X.~L. {De{\'a}n-Ben}, A.~Cossettini, M.~Magno, D.~Razansky, and L.~Benini, ``{{FPGA-Accelerated Hybrid Lossless}} and {{Lossy Compression}} for {{Next-Generation Portable Optoacoustic Platforms}},'' in \emph{2024 {{IEEE Ultrasonics}}, {{Ferroelectrics}}, and {{Frequency Control Joint Symposium}} ({{UFFC-JS}})}, Sep. 2024, pp. 1--5.

\bibitem{cossettiniRDMAInterfaceUltraFast2022}
A.~Cossettini, K.~Taranov, C.~Vogt, M.~Magno, T.~Hoefler, and L.~Benini, ``A {{RDMA Interface}} for {{Ultra-Fast Ultrasound Data-Streaming}} over an {{Optical Link}},'' in \emph{2022 {{Design}}, {{Automation}} \& {{Test}} in {{Europe Conference}} \& {{Exhibition}} ({{DATE}})}, Mar. 2022, pp. 80--83.

\bibitem{villaniAdaptiveImageReconstruction2024}
F.~Villani, {\c C}.~{\"O}zsoy, C.~Vogt, A.~Cossettini, X.~L. {De{\'a}n-Ben}, M.~Magno, D.~Razansky, and L.~Benini, ``Adaptive {{Image Reconstruction}} for {{Optoacoustic Tomography}}: {{A Partial FPGA Reconfiguration Approach}},'' \emph{IEEE Sensors Letters}, vol.~8, no.~8, pp. 1--4, Aug. 2024.

\end{thebibliography}

\end{document}